\documentclass[prb,showpacs,twocolumn,floatfix,amsmath]{revtex4}
\usepackage{graphicx} \usepackage{dcolumn} \usepackage{bm}
\usepackage{float}
\usepackage{marvosym,textcomp}

\begin{document}
\title{Three dimensional filamentary structures of a relativistic electron beam
in Fast Ignition plasmas}
\author{Anupam Karmakar}\email{anupam@tp1.uni-duesseldorf.de}
\author{Naveen Kumar}
\author{Alexander Pukhov}
\affiliation{Institut f\"ur Theoretische Physik I, Heinrich-Heine-Universit\"at,
D\"usseldorf, 40225, Germany}
\author{Oleg Polomarov}
\author{Gennady Shvets}
\affiliation{Department of Physics and Institute for Fusion Studies, University
of Texas at Austin, One University Station, Austin, TX 78712, USA}

\begin{abstract}
The filamentary structures and associated electromagnetic fields of a
relativistic electron beam have been studied by three dimensional
particle-in-cell (PIC) simulations in the context of Fast Ignition fusion. The
simulations explicitly include collisions in return plasma current and
distinctly examine the effects of beam temperature and collisions on
the growth of filamentary structures generated.
\end{abstract}

\maketitle

The transport of laser generated electron beams in overdense plasmas
is an important topic to study specially in the context of Fast Ignition
(FI)\cite{tabak94}. PIC simulations and calculations have shown that
the typical energy of electrons in these beams is of the order of a
few MeV, which implies Giga-Ampere currents needed to transport
Peta-Watt energy fluxes \cite{pukhov96,sentoku03}. However, the
transportation of beam currents greater than the Alfv\'en current limit is not
possible as the self-generated magnetic field associated with these currents
bends electron trajectories strongly thereby preventing the forward
propagation of the beam. In vacuum, the Alfv\'en current limit is
$I_{alf}=17 \gamma\beta \,k$A, where $\gamma=1/\sqrt{1-\beta^2}$ is
the relativistic Lorentz factor associated with the beam. A current
greater than this limit can only propagate as a charge neutralized current. This
sort of situation occurs in the FI scenario, where the charge of the
relativistic electron beam is compensated by the return plasma current and the
propagation of the electron beam is possible. But this configuration, where two
opposite streams of plasma currents are counterpropagating, is susceptible to
the electrostatic two-stream instability and the transverse filamentation
instability, often referred as the Weibel instability. Due to these
instabilities, the electron beam gets splitted into many filaments
\cite{lampe73,honda2000, tabak08}. This filamentation can significantly affect
the beam energy deposition at the center of the fuel target and thus is one of
the central issues for the FI concept. Recently, Honrubia \emph{et al.}
\cite{honrubia06} have carried out 3D hybrid simulations of the fast electron
transport and resistive beam filamentation in inertial fusion plasma. Califano
\emph{et al.}~\cite{califano06} have reported three dimensional magnetic
structures generated due to the Weibel instability of a relativistic electron
beam in collisionless plasmas. Three dimensional PIC simulations of the Weibel
instability in astrophysical context have also been reported \cite{fonseca03}.
Moreover, the evidence of Weibel-like dynamics and the resultant filamentation
of electron beams have been observed experimentally~\cite{jung05}.

The role of collisions in the Weibel instability is rather non-trivial and has
been studied analytically in linear and quasilinear regime of the instability
\cite{deutsch05}. The effect of collisions on the nonlinear regime of the
instability is rather unaccessible to analytical calculations and can best be
studied by PIC simulations. However, till now the full scale PIC simulations on
the filamentation of the electron beams  have mostly concentrated 
on the collisionless Weibel instability and have not explicitly included the
effect of the collisions in the return plasma current. At the same
time, collisions in the return plasma current play an important role on the
Weibel instability. The eigenmodes of a collisional plasma have been computed
recently, where it was shown that frictional nature of collisions can change the
dynamics of the eigenmodes of the system significantly \cite{honda04}. In the
presence of collisions the Weibel instability can not be suppressed by the
transverse beam temperature as proposed initially \cite{silva02}.  A more
focused discussion about the collisional effects on the Weibel instability has
also been addressed recently where it was shown that even small collisions in
the background plasma can trigger the Weibel instability of a warm electron beam
that would be otherwise stable in collisionless plasma
\cite{karmakar08}.

Motivated by these findings, we carry out three-dimensional PIC simulations on
the transport of a relativistic electron beam in the FI plasmas by explicitly
including the effect of collisions in the return plasma current. The
rationale behind choosing the 3D geometry is twofold: the first reason is to
study the realistic geometry and the second is to show that in this geometry,
the Weibel instability can't be suppressed due to the transverse beam
temperature. It has been shown before analytically that in the 3D
geometry the Weibel instability survives even with high transverse beam
temperature due to the presence of an ``oblique mode'', which is the
manifestation of coupling between the Weibel instability and the two-stream
instability \cite{bret05}. We conjecture the existence of the ``oblique mode''
due to the anomalous collisionality of the background plasma. It arises
due to the turbulence of electrostatic fields generated by the two-stream
instability. We choose four cases for the simulations (A) cold electron beam in
a collisionless background plasma, (B) cold electron beam in a collisional
background plasma, (C) warm electron beam in a cold collisionless background
plasma, and (D) warm electron beam in a collisional background plasma, thus
highlighting the influence of these physical processes separately. Hence, for
the first time we have studied the filamentation of the relativistic beam
investigating the role of each physical process clearly in  three
dimensional particle-in-cell simulations. 

In our simulations, the electron beam propagates along the negative
$Z-$axis with the relativistic velocity $\upsilon_{(b,z)} \gg
\upsilon_{(b,x)}, \upsilon_{(b,y)}$. The bulk cold background plasma is
represented by ambient plasma electrons, while the plasma ions are considered as
a fixed charge-neutralizing background with the density $n_0=n_b + n_p$, where
$n_b$ and $n_p$ are the densities of the electron beam and the
background plasma, respectively. Initially the beam current is fully compensated
by the return plasma current. The beam density is much smaller than the
background plasma electron density, {\it i.e.} $n_{b} \gg n_{p}$, a usual
scenario of the FI. The spatial dimensions of the simulation domain $L$ is large
in comparison with the electron skin depth \emph{i.e.} $L \gg \lambda_{s}$,
where $\lambda_{s}=c/\omega_{pe}$; $c$ and $\omega_{pe}$ are the velocity of
light in vacuum and the total electron plasma frequency, respectively. The
quasi-neutrality is maintained over all simulation time as the field evolutions
due to the Weibel instability occur on a time scale slower than the plasma
electron frequency $\Delta t>>1/\omega_{pe}$. The collisional processes are
simulated with a newly implemented collision module in the relativistic PIC
code Virtual Laser Plasma Laboratory (VLPL)~\cite{pukhov99}.

The 3D simulation box has spatial dimensions $X \times Y \times Z = \left(20
\lambda_s \times20 \lambda_s \times20 \lambda_s\right)$. The 3D simulation
domain is sampled with the mesh of $160 \times 80 \times 20$ cells. All
simulations are performed with 64 particles per cell and with a grid size much
smaller than the background plasma skin depth ($h_x, h_y, h_z << \lambda_{s}$).
The density ratio between the beam and plasma electrons is $n_{p}/n_{b}=9$,
whereas the beam and the background plasma electrons have velocities
$\upsilon_b=0.9\,c$ and $\upsilon_p=0.1\,c$. The evolution of the
field energies for every component $F_i$ of the fields $\mathbf{E}$
and $\mathbf{B}$ is recorded at every diagnostic step of the
simulation summed over all the grid cells as 

\begin{equation}\label{eq:3dfields} 
\int_V F_i^2~dxdydz=\sum{\left(eF_i m_e c\,\omega_{pe}\right)^2 h_x
h_y h_z},
\end{equation}

\noindent where, $h_x$, $h_y$ and $h_z$ are the spatial grid sizes along $x$,
$y$ and $z$ and ${eF_i}/{m_ec\omega_{pe}}$ represents the relativistic field
normalizations. The relativistic electron beam has a temperature of
$T_{b}\approx 70$ keV  and the ambient plasma collision frequency is
assumed to be $\nu_{ei}/\omega_{pe}=0.15$ for these simulations. The
background plasma is always cold initially while the beam electrons do 
not face any collisions.

\begin{figure}[floatfix]
\centering
\includegraphics[width=0.45\textwidth,keepaspectratio]
{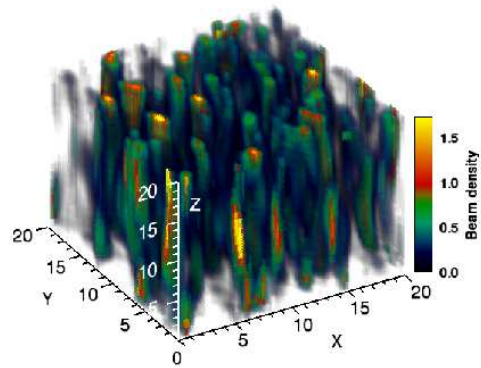}
\caption{(color online) 3D structure of the beam filaments generated in the PIC
simulation at a time $T=16(2\pi/\omega_{pe})$ in the nonlinear regime for the
case (A), where the electron beam is cold and the plasma is collisionless. The
filaments are tiny and their transverse dimension is comparable to the skin
length of the cold background plasma.}
\label{fig:collisionless}
\end{figure}

\begin{figure}[floatfix]
\centering
\includegraphics[width=0.45\textwidth,keepaspectratio]{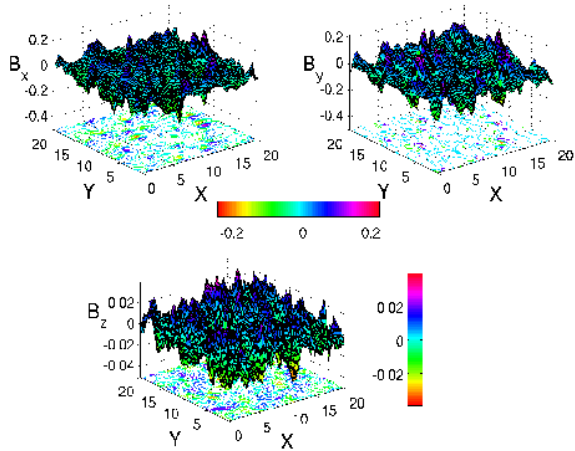}
\caption{(color online)~Illustration of the transverse ($B_x$, $B_y$) and
longitudinal ($B_z$) magnetic fields at the same time as
Fig.~\ref{fig:collisionless}. The contour lines on the bottom surface
gives a 2D projection. {\bf B} fields are normalized as
explained in the 
Eq.~\eqref{eq:3dfields}. Magnitude of the transverse components are larger
almost by an order than the longitudinal component.}
\label{fig:b_fields}
\end{figure}

\begin{figure}[floatfix]
\centering
\includegraphics[width=0.45\textwidth,keepaspectratio]{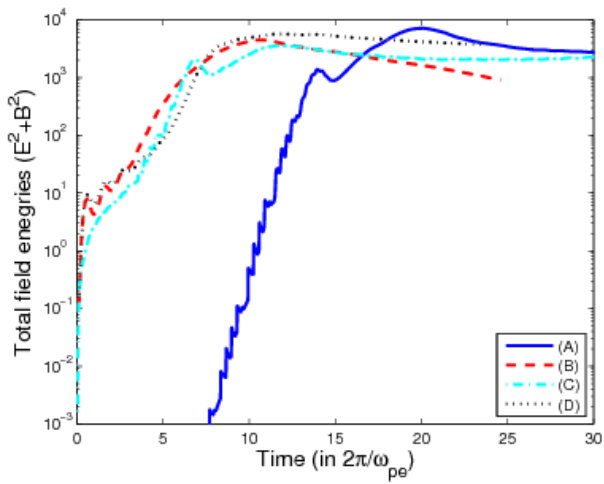}
\caption{(color online)~Time evolution of Weibel field energies ($E^2$ and
$B^2$) for the different simulation cases: (A) cold electron beam in a
collisionless background plasma, (B) cold electron beam in a collisional
background plasma, (C) hot electron beam in a collisionless plasma and (D) hot
electron beam in collisional background plasma. The time is normalized in units
of $2\pi/\omega_{pe}$ and the $\mathbf{E}$ and $\mathbf{B}$ fields are
normalized as $(e\mathbf{E}/m_e\omega_{pe}c)$ and
($e\mathbf{B}/m_e\omega_{pe}c$), respectively.}
\label{fig:growthrates}
\end{figure}

Fig.~\ref{fig:collisionless} shows the structure of the beam filaments during
the nonlinear regime at time $T=16(2\pi/\omega_{pe})$ for the case (A). The
filaments transverse spatial sizes are comparable to the plasma skin depth
$\lambda_{s}$ while their longitudinal spatial extent is larger than the plasma
skin depth. This could be understood as follows: in the linear regime of the
beam filamentation, the transverse extent of the filaments is governed by the
Weibel instability and its growth rate maximizes around the plasma skin depth
in $k$-space. However, the longitudinal extent is determined by the two-stream
instability which maximizes at wavelengths much larger than the plasma
skin depth. As a result, the filaments have small spatial extent in
the transverse direction and elongated in the propagation direction.
Later on, during the nonlinear stage of the instability, the filaments merge
together and their radii grow. The growth rate of the filamentation instability
depends on the beam-to-plasma densities ratio, and is expressed as
$\Gamma=(\sqrt{3}/2^{4/3})\left(n_b/n_p\gamma_b\right)^{1/3}$, where
$\gamma_b$ is the relativistic Lorentz factor of the electron beam
\cite{fainberg70}. The Weibel instability generates very strong quasi-static
magnetic field, which can be seen from the surface plot of the transverse
($B_x$, $B_y$) and longitudinal ($B_z$) magnetic field $B_x$ shown in
Fig.~\ref{fig:b_fields}. Evidently, the amplitude of the transverse components
is larger at least by one order of magnitude compared to the longitudinal
component. The contour lines on the bottom surface in this figure show that each
of these tiny filaments is surrounded by strong axial magnetic fields. We have
plotted the temporal growth of such field energies for all four simulation cases
later in Fig.~\ref{fig:growthrates}. 

\begin{figure}[floatfix]
\centering
\includegraphics[width=0.45\textwidth]{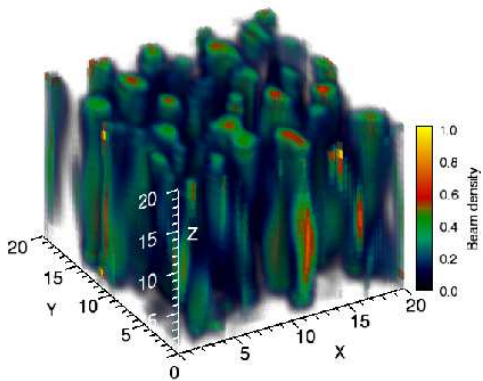}
\caption{(color online)~ 3D structure of the beam filaments from the PIC
simulation in a time $T=8(2\pi/\omega_{pe})$ for case (B), cold
electron beam in collisional plasma. The filament sizes are larger than
collisionless case due to collisional dissipation.}
\label{fig:cold_collision}
\end{figure}

The simulation case (B) is shown in Fig.~\ref{fig:cold_collision}. We observe
that the transverse spatial width of the filaments is larger while the
longitudinal extent remains unchanged. Due to collisions, the growth rate of the
Weibel instability saturates at large plasma skin depth in $k$-space. At later
times, the filaments merge together and their transverse spatial extent grows
even larger, which is evident from the simulation results. 

\begin{figure}[floatfix]
\centering
\includegraphics[width=0.45\textwidth,keepaspectratio]{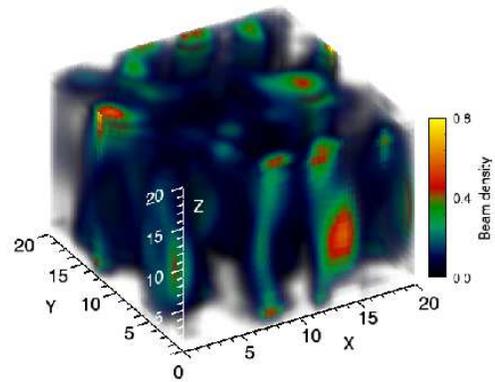}
\caption{(color online) Beam filamentation persists in the case (C), where the
electron beam is warm with $T_b\approx 70keV$ propagating in a collisionless
plasma, at time $T=10(2\pi/\omega_{pe})$ in the nonlinear regime. The evacuation
of the background plasma is not complete due the strong thermal effects.}
\label{fig:temp_collisionless}
\end{figure}

The filament structures of the simulation case (C), where the electron
beam has a transverse beam temperature $T_b \approx 70$ keV,  is shown in
Fig.~\ref{fig:temp_collisionless} at time $T=10(2\pi/\omega_{pe})$. The Weibel
instability could be suppressed by the high transverse beam temperature.
However, in $3$D geometry the coupling of the Weibel and the two-stream
instabilities can manifests itself into the so called oblique mode, which
persists even when the beam temperature is high. We may also consider the
occurrence of the beam filamentation in this case due to ``anomalous plasma
collisionality'' as shown in Ref.\cite{karmakar08}. The two-stream
mode actually generates electrostatic turbulence in the plasma. Stochastic
fields associated with this turbulence scatter the beam and plasma electrons and
lead to an effective collisionality in the return current. This anomalous effect
revives back the Weibel instability. The phenomena of anomalous resistivity of
plasma generated due to the turbulence of electric and magnetic fields have been
noted earlier also \cite{sentoku03}. The allowance for finite background plasma
temperature decreases the growth rate of the instability. The interplay of
collisions in different regimes of beam and background plasma temperatures on
the Weibel instability can be found in Ref.~\cite{deutsch05}. Moreover,
it maybe worthwhile to mention here that this filamentation is lacking complete
evacuation of the background plasma electrons contrary to the previous cases,
presumably due to the strong thermal effects.

\begin{figure}[floatfix]
\centering
\includegraphics[width=0.45\textwidth,keepaspectratio]
{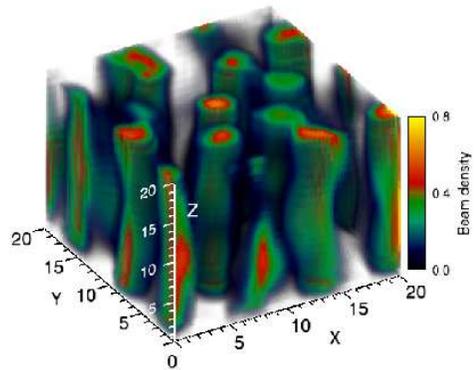}
\caption{Structure of the electron beam filaments derived from the PIC
simulations for the case (D), where the warm electron beam is propagating in the
collisional background plasma. The filaments are larger and total evacuation of
the background plasma is prominent.}
\label{fig:warm_collision}
\end{figure}

Fig.~\ref{fig:warm_collision} shows the beam filaments of the last simulation
case (D), where the electron beam is hot and the background plasma is
collisional. Here again we see strong filamentation and the spatial width of
the beam filaments are also larger due to collective collisional plasma
effects. Moreover, we have observed almost complete evacuation of the
background plasma in this case.

To get further insight into the growth of the fields associated with the
filamentation instability, we have studied the the evolution of electric and
magnetic field energies, shown in Fig.~\ref{fig:growthrates}, for the four
different simulation cases. The vertical axes in Fig.~\ref{fig:growthrates}
represents the normalized field energies in logarithmic scales whereas the
horizontal axes are for time scaled in $2\pi/\omega_{pe}$. In each of these
cases, we see a stage of linear instability, where the field energies build up
exponentially in time and it is followed by the nonlinear saturation of the
instability. Afterwords, in the nonlinear stage of the instability, the
filaments merge rapidly with each other due to the magnetic attraction and the
field energies saturate.  One may note that the simulation cases (B), (C), and
(D) display similar trend of field energy build up. This similarity is very
interesting for the case (C), where the electron beam is warm and has
the sufficiently high temperature to kill the Weibel instability. Yet, we still
see the filamentation. Clearly it happens due to the presence of the two-stream
or ``oblique mode''. Based on the similarity between cases (B) and (C), we
attribute the filamentation of the beam to the `effective plasma
collisionality' of the beam system as discussed before in the
manuscript as well \cite{karmakar08}.

In summary, we have studied the filamentation instability of a relativistic
electron beam in Fast Ignition plasma by  three dimensional particle-in-cell
simulations, distinctly examining the effects of high beam temperature
and plasma collisions. The important result of this study is that the
beam temperature does not suppress the filamentation instability even in
absence of collisions in beam plasma system. An explanation on the
persistence of the Weibel instability in 3D geometry is offered. It is
attributed to the anomalous collisionality of the beam-plasma system due to the
two-stream mode. The nonlinear regime of the filaments merging and the growth of
field energies due to plasma collisions have been studied.

This work was supported by the DFG through TR-18 project and the US Department
of Energy grants DE-FG02-04ER41321 and DE-FG02-04ER54763.

\end{document}